\documentclass{PoS}

\usepackage[latin1]{inputenc}
\usepackage{amsmath}
\usepackage{graphicx}
\usepackage{verbatim}

\newcommand{\bc}{\begin{center}}
\newcommand{\ec}{\end{center}}
\newcommand{\be}{\begin{equation}}
\newcommand{\ee}{\end{equation}}
\newcommand{\bea}{\begin{eqnarray}}
\newcommand{\eea}{\end{eqnarray}}
\newcommand{\beqn}{\begin{eqnarray}}
\newcommand{\eeqn}{\end{eqnarray}}
\newcommand{\ba}{\begin{array}}
\newcommand{\ea}{\end{array}}
\newcommand{\bal}{\begin{aligned}}
\newcommand{\eal}{\end{aligned}}
\newcommand{\ben}{\begin{enumerate}}
\newcommand{\een}{\end{enumerate}}
\newcommand{\bitem}{\begin{itemize}}
\newcommand{\eitem}{\end{itemize}}

\newcommand{\fr}{\frac}

\newcommand{\crn}{\nonumber \\}

\newcommand{\noi}{\noindent}

\newcommand{\OO}{\mathcal{O}}

\newcommand{\eq}[1]{Eq.~(\ref{#1})}

\newcommand{\fig}[1]{Fig.~\ref{#1}}

\newcommand{\gev}{{\unskip\,\text{GeV}}}
\newcommand{\tev}{{\unskip\,\text{TeV}}}

\newcommand{\fb}{{\unskip\,\text{fb}}}

\newcommand{\hour}{{\text{h}}}

\newcommand{\fc}{{\texttt{FormCalc-6.0}}}
\newcommand{\fa}{{\texttt{FeynArts}}}
\newcommand{\sloops}{{\texttt{SloopS}}}

\newcommand{\epem}{e^+e^-}

\newcommand{\eezzz}{e^+e^-\to ZZZ}
\newcommand{\eezzzt}{$e^+e^-\to ZZZ\;$}
\newcommand{\eezzztp}{$e^+e^-\to ZZZ$}
\newcommand{\eezzzgam}{e^+e^-\to ZZZ\gamma}
\newcommand{\eewwz}{e^+e^-\to W^+W^- Z}
\newcommand{\eewwzt}{$e^+e^-\to W^+W^- Z \;$}

\newcommand{\eewwzgam}{e^+e^-\to W^+W^- Z\gamma}
\newcommand{\eewwt}{$e^+e^-\to W^+W^-  \;$}

\newcommand{\eewwgam}{e^+e^-\to W^+W^- \gamma}

\newcommand{\eeww}{{$e^+e^-\to W^+W^-$}}

\def\slashepi{\epsilon_i\kern -.720em {/}}
\def\slashpi{p_i\kern -.600em {/}}

\vspace*{0.1cm}\rightline{MPP-2010-176}

\title{NLO corrections to WWZ and ZZZ production at the ILC}

\ShortTitle{NLO corrections to WWZ and ZZZ production at the ILC}

\author{Fawzi Boudjema \\
LAPTH, Universit{\'e} de Savoie, CNRS, \\
BP110, F-74941 Annecy-le-Vieux Cedex, France
\\ 
E-mail: \email{boudjema@lapp.in2p3.fr}
      }

\author{\speaker{Le Duc Ninh}, Marcus M. Weber \\
       Max-Planck-Institut f{\"u}r Physik (Werner-Heisenberg-Institut),\\
       D-80805M{\"u}nchen, Germany\\
       E-mail: \email{leducninh@gmail.com}, \email{mmweber@mppmu.mpg.de}}
      
\author{Sun Hao  \\
Department of Physics, State University of NewYork, Buffalo, NY14260, USA
\\ 
E-mail: \email{hsun6@buffalo.edu}
      }

\abstract{We calculate the full one-loop electroweak corrections to tri-boson
production (ZZZ and WWZ) at the ILC. This is important to understand
the Standard Model (SM) gauge quartic couplings which can be a window
on the mechanism of spontaneous symmetry breaking. We find that even
after subtracting the leading QED corrections, the electroweak
corrections can still be large especially as the energy increases.}

\FullConference{3rd Computational Particle Physics Workshop\\
                September 23-25, 2010 \\
                KEK Tsukuba Japan}

\begin{document}
\section{Introduction} 
Due to its clean environment an $e^+e^-$ linear collider 
in the TeV range is an ideal machine to probe 
in detail and with precision the inner working of 
the electroweak structure, in particular the mechansim of symmetry breaking.
From this perspective the
study of \eewwzt and \eezzzt may be very instructive and would
play a role similar to $\epem \to W^+W^-$ at lower energies.
Indeed it has been stressed  that  \eewwzt
and \eezzzt  are prime processes for probing the quartic vector
boson couplings \cite{Belanger:1992qh}. In particular deviations from the gauge value in
the quartic $W^+W^-ZZ$ and $ZZZZ$ couplings that are accessible in
these reactions might be the residual effect of physics intimately
related to electroweak symmetry breaking. Since these effects
can be small and subtle, knowing these cross sections with high
precision is mandatory. This calls for theoretical predictions
taking into account loop corrections.

In this report we discuss the next-to-leading order (NLO) corrections to the processes \eezzzt 
and \eewwzt at the future international linear collider (ILC). We also address some technical issues 
related to numerically stable evaluation of one-loop integrals, which is a challenge for one-loop multi-leg 
automatic calculations. 

\section{Tree level}
\label{sect-tree}

\begin{figure}[h]
\begin{center}
\includegraphics[width=0.7\textwidth]{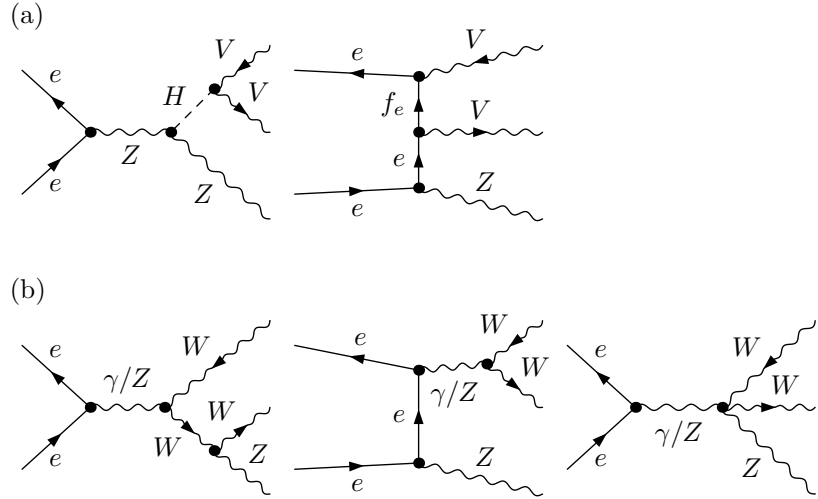}
\end{center}
  \caption{{\em Representative Born diagrams for $\eezzz$ and $\eewwz$. Diagrams (a) contribute to
    both processes while  diagrams of type (b) contribute only to $\eewwz$. The first diagram of type (a)
    will be referred to as the Higgsstrahlung contribution.}}
  \label{fig:treediag}
\end{figure}

At leading order $W^+W^-Z$ and $ZZZ$ final states are produced
through the diagrams shown in \fig{fig:treediag}. The important gauge 
couplings fermion-fermion-vector, $\gamma WW$ and $ZWW$, which also 
appear in the well-tested \eeww process, have been measured and found 
in good agreement with the prediction of the SM \cite{Amsler:2008zzb, LEPworking}.
Both processes include the Higgsstrahlung contribution where the
splitting $H^\star \to VV$ occurs. This contribution is small and vanishes in the large Higgs mass limit. 
Since the precision electroweak data
suggest a Higgs mass below the $WW$ threshold, we restrict our study to the
region $M_H < 160\gev$.
This means that the Higgsstrahlung
contribution can not be resonant and therefore in our
calculation no width is introduced. The important features at tree level are that the neutral quartic 
gauge coupling $ZZZZ$ vanishes in the SM and the charged $\gamma ZWW$, $ZZWW$ couplings 
occur in the $\eewwz$ process. In practice, the coupling $\gamma ZWW$ (and also $\gamma\gamma WW$) can 
be tested at lower energies via the process $\eewwgam$.

\section{NLO calculations}
\label{sect-nlo}

Our calculations are done in the framework of the SM. 
The virtual corrections have been evaluated using a conventional
Feynman-diagram based approach using standard techniques of tensor 
reduction. 
We use the packages \fa\ and \fc\ to generate all Feynman diagrams 
and helicity amplitude expressions \cite{fafc}. 
We also use \sloops\ to check the 
correctness of the amplitudes by checking non-linear gauge invariance 
(see \cite{sloops} and references therein). 
The total number of diagrams in the 't~Hooft-Feynman
gauge is about 2700
including 109 pentagon diagrams for $\eewwz$ and about 1800
including 64 pentagons for $\eezzz$. This already shows that
$\eewwz$ with as many as 109 pentagons is more challenging than
$\eezzz$. 
Indeed getting stable results for all scalar and tensor (up to rank 4) box integrals in 
the process $\eewwz$ is a highly nontrivial task. The five-point integrals are reduced to 
four-point integrals by using the method of Denner-Dittmaier \cite{Denner:2002ii} which 
does not involve the Gram determinant, $\det G = \det (2p_i\cdot p_j)$ with $p_{i,j}$ being the 
external momenta, in the denominator. The four-point (and three-point) tensor integrals are in turn 
recursively reduced to scalar integrals by using Passarino-Veltman (PV) method. The problem 
with this method is that the numerical results become unstable when the Gram determinant is small. 
For instance, the result for a rank-4 box integral includes tensor coefficients of the form
\bea
D_{ijkl} = \fr{N(p,m)}{(\det G)^4},
\label{eq_gram}
\eea
where the numerator is a complicated function of internal masses and external momenta whose indices have been excluded for simplicity. In many cases the function $N$ vanishes in the limit $\det G\to 0$, leaving the tensor coefficients finite. In particular, this is usually true if the internal particles are
massive. This non-trivial behavior of the numerator which is a linear combination of scalar integrals can 
be spoiled by numerical cancellation or inconsistent approximations (like small mass/momentum approximations) in calculation of the 
scalar integrals, leading to numerical instabilities in the right-hand side of \eq{eq_gram} when $\det G$ becomes small. A good way to solve this problem is therefore using higher-precision arithmetic in the calculation of loop integrals when numerical cancellation occurs.

It is important to notice that the PV method fails when $\det G$ is exactly zero. If this happens 
the $N$-point 
function of rank $M$ can be written as a combination of $(N 
- 1)$-point functions of rank $M$. 
This is called segmentation \cite{boudjema_temes}. We have exploited this fact to avoid the small $\det G$ region by using segmentation if the following condition is met
\begin{eqnarray}
\label{cdtgram} \fr{\det(G)}{(2p_{max}^2)^3}<10^{-7},
\end{eqnarray}
where $p_{max}^2$ is the maximum external mass of a box diagram. This extrapolation is used only for the four-point integrals and turns out to be good enough for the present calculations. We have compared this to the method of using higher precision arithmetic (quadruple precision for the Fortran 77 code) and 
obtained good agreement. 

\begin{figure}[t]
  \centering
  \includegraphics[width=0.6\textwidth]{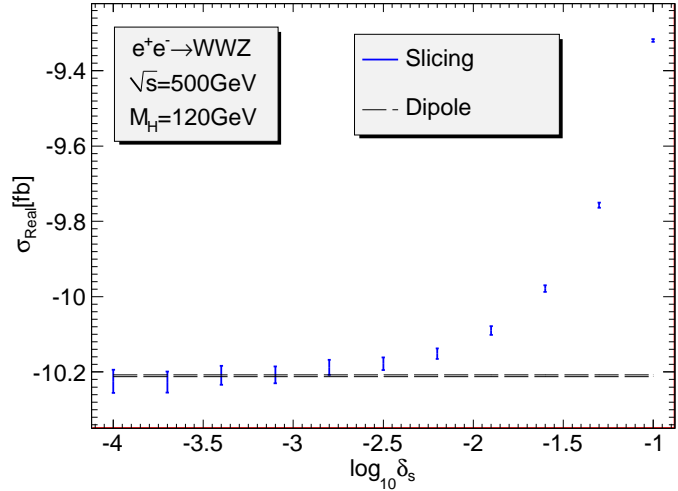}
  \caption{{\em Dependence of $\sigma_{\text{real}}^{\eewwzgam}$ on the soft
      cutoff $\delta_s$ ($E_\gamma < \delta_s\sqrt{s}/2$) in phase-space slicing. Only the non-singular part is shown, i.e. the IR singular
      $\ln(m_\gamma^2)$ terms are set to zero. The result using dipole
      subtraction is shown for comparison with the error given by the width
      of the band.}}
  \label{fig:dipslicmp}
\end{figure}
In addition to the virtual corrections we also have to consider
real photon emission, {\it i.e.} the processes $\eewwzgam$ and
$\eezzzgam$. The corresponding amplitudes are divergent in the
soft and collinear limits. The soft singularities cancel against
the ones in the virtual corrections while the collinear
singularities are regularized by the physical electron mass.  To
extract the singularities from the real corrections and combine
them with the virtual contribution we apply both the dipole
subtraction scheme and a phase space slicing method. 
The former is used to produce the final results since it yields smaller 
integration errors as shown in \fig{fig:dipslicmp}. 
Further details are given in \cite{Boudjema:2009pw}.

It is well-known that the collinear QED correction related to
initial state radiation in $e^{+}e^{-}$-processes is large.
In order to see the effect of
the weak corrections, one should separate this large QED correction from
the full result. It means that we can define the weak correction as
an infrared and collinear finite quantity. The definition we adopt
in this paper is based on the dipole subtraction formalism. In this
approach, the sum of the virtual and the so-called "endpoint" 
(see \cite{Dittmaier:1999mb} for the definition) contributions satisfies
the above conditions and can be chosen as a definition for the weak
correction
\beqn
\sigma_\text{weak} = \sigma_\text{virt} + \sigma_\text{endpoint}.
\eeqn
For the numerical results shown in the next section,
we will make use of this definition.

Before presenting our numerical results it is stressed that 
we have performed the calculation in at least two independent ways
both for the virtual and the real corrections leading to two
independent numerical codes (one code is written in Fortran 77, the other in C++). 
A comparison of both codes has shown
full agreement at the level of the integrated  cross
sections as well as all the distributions that we have studied. 
Moreover, we have done detailed comparisons with other groups \cite{JiJuan:2008nn, Wei:2009hq} 
and obtained good agreement.

\section{Numerical results}
To absorb large corrections to the electromagnetic coupling and universal corrections 
due to the isospin breaking effects we use 
\bea
\alpha &=& \alpha_{G_\mu}=\fr{\sqrt{2}G_\mu M_W^2}{\pi} \sin^2\theta_W,\crn
\alpha_{G_\mu} &=& \alpha(0)(1 + \Delta r)
\eea 
with $G_\mu$ denoting the Fermi constant and $\theta_W$ is the weak-mixing angle, at tree level. The explicit 
form of $\Delta r$ at one-loop order together with all the input parameters are given in \cite{Boudjema:2009pw}. When calculating the NLO corrections we have to subtract the one-loop $\Delta r$ contribution to avoid double counting. Since the real photon corrections are proportional to $\alpha(0)$, we require 
the full NLO corrections to be of order $\OO(\alpha_{G_\mu}^3\alpha(0))$. 

\label{sect-results}
\begin{figure}[htb]
\begin{center}
\mbox{\includegraphics[width=0.45\textwidth
]{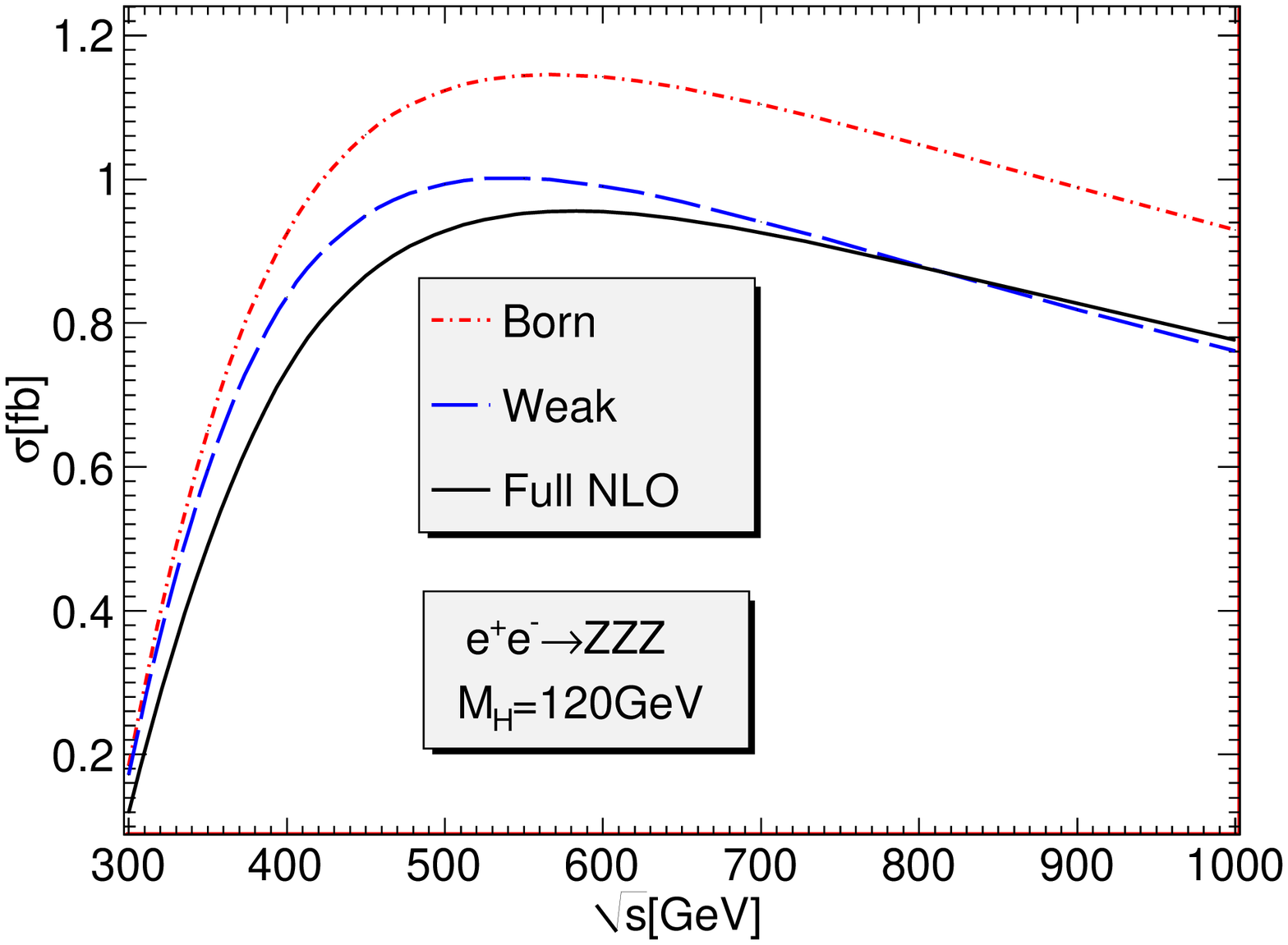} \hspace*{0.01\textwidth}
\includegraphics[width=0.45\textwidth]{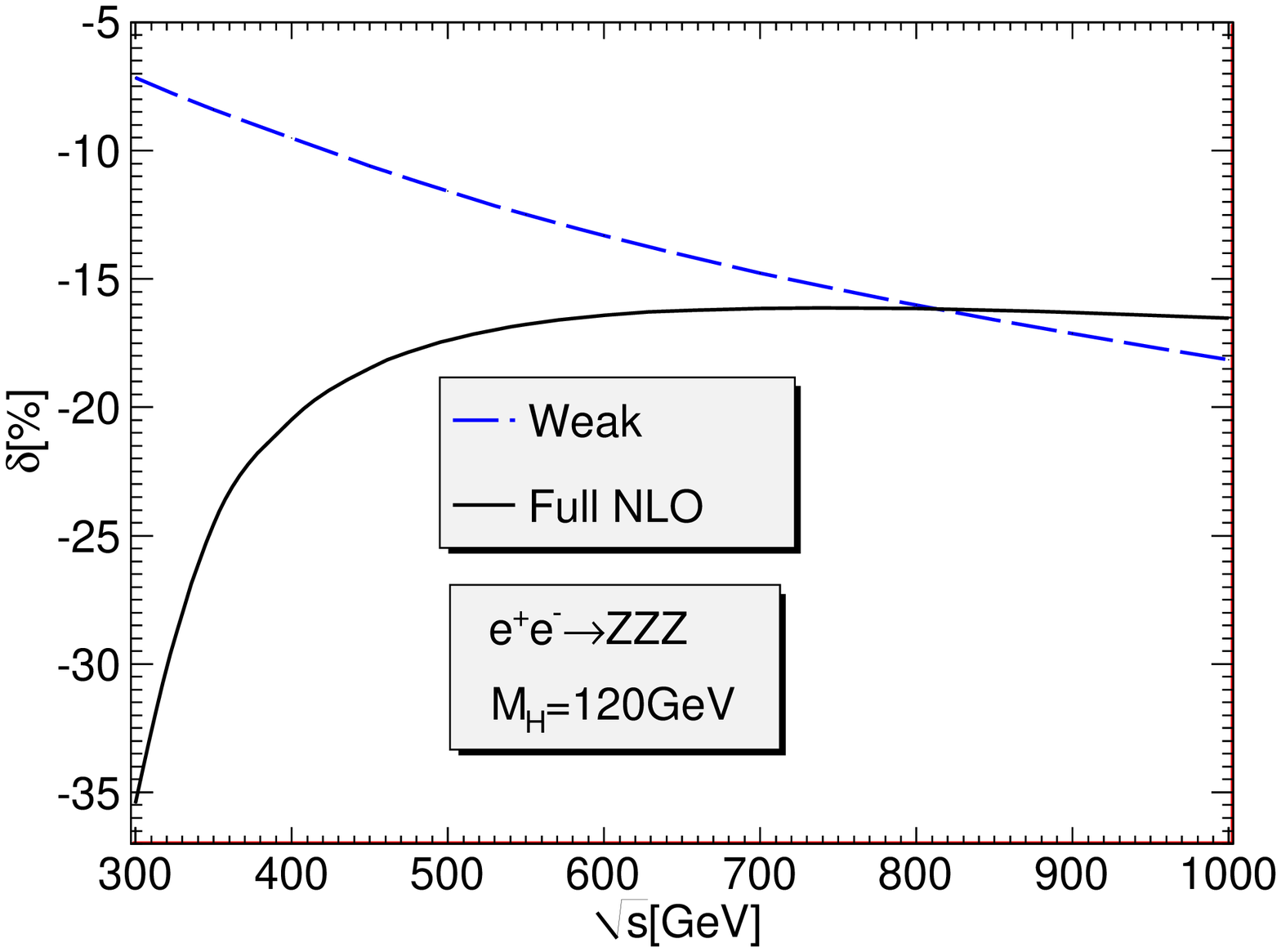}}
\mbox{\includegraphics[width=0.45\textwidth
]{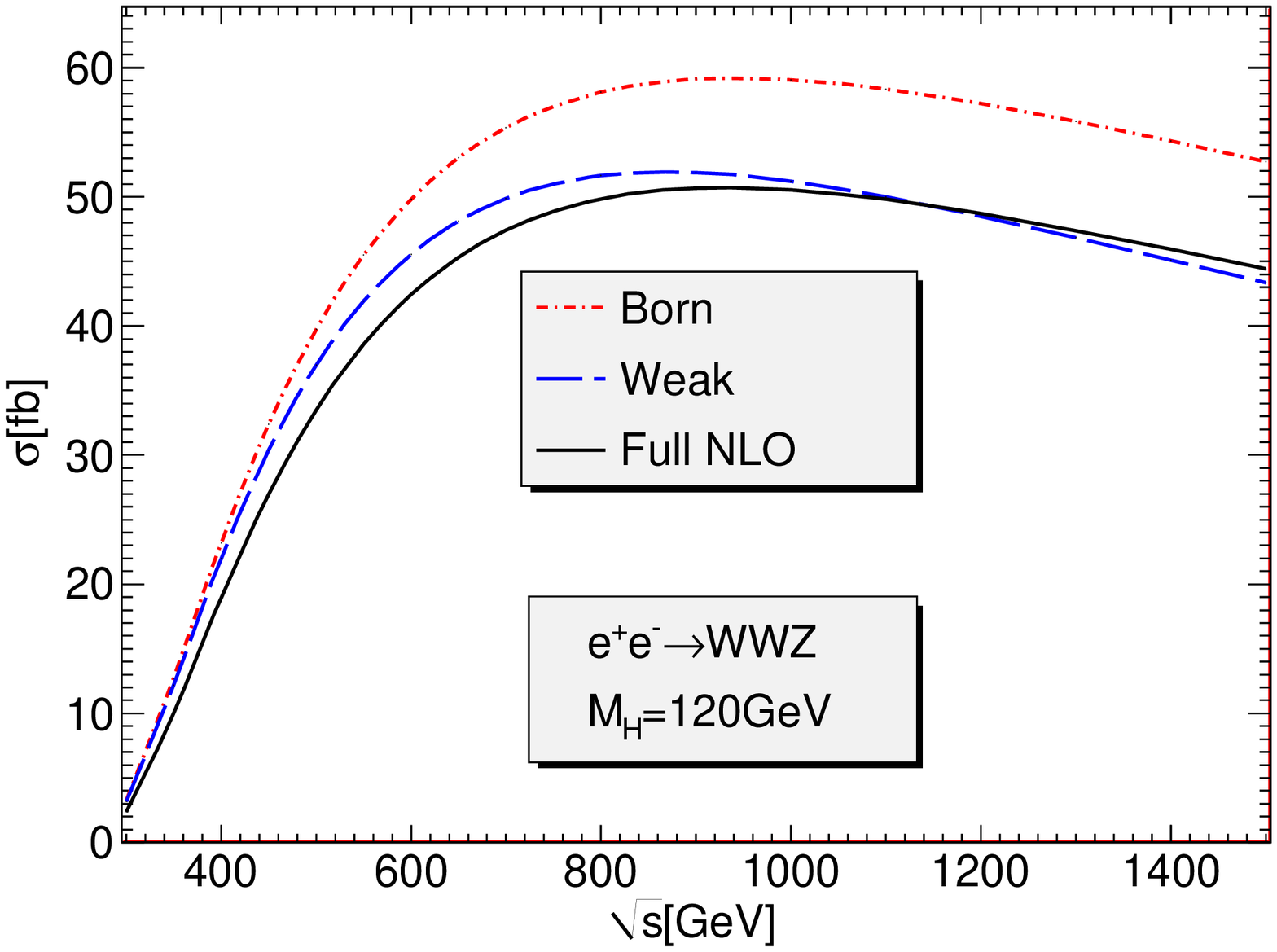} \hspace*{0.01\textwidth}
\includegraphics[width=0.45\textwidth]{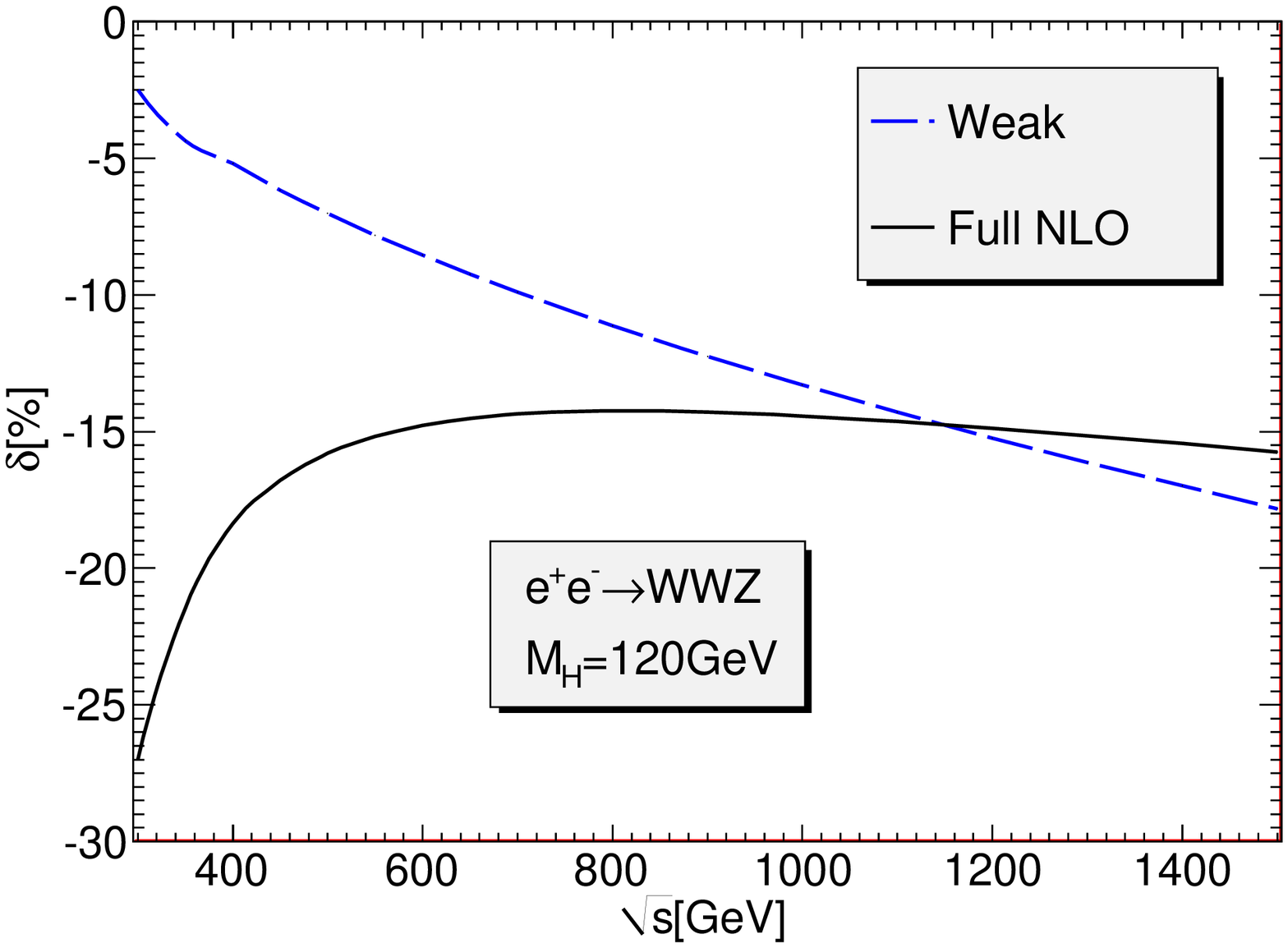}}
\caption{\label{ourvvzsigma}{\em The total cross section
for $\eezzz$ (top) and $\eewwz$ (bottom) 
as a function of $\sqrt{s}$ for the Born, full ${\cal
O}(\alpha)$
 and genuine weak correction.
The panels on
the left show the Born, the full NLO and the weak correction.
The panels on the right show the
corresponding relative (to the Born) percentage corrections.
}}
\end{center}
\end{figure}

\noindent {\underline{$\eezzz$:}}\\
As shown in Fig.~\ref{ourvvzsigma} the tree-level cross section
rises sharply once the threshold for production opens, reaches a
peak of about $1.1\fb$ around a centre-of-mass energy of $600\gev$
before very slowly decreasing with a value of about $0.9\fb$ at
1\tev. The full NLO
corrections are quite
large and negative around threshold, $-35\%$, decreasing sharply to stabilise
at a plateau around $\sqrt{s}=600\gev$ with $-16\%$ correction. The
sharp rise and negative corrections at low energies are easily
understood. They are essentially due to initial state radiation
(ISR) and the behaviour of the tree-level cross section. The
photon radiation reduces the effective
centre-of-mass energy and therefore explains what is observed in the
figure.
On the other hand the genuine weak corrections, in the
$G_\mu$ scheme, are relatively small at threshold, $-7\%$. They
however  increase steadily with a correction as large as $-18\%$
at $\sqrt{s}=1\tev$.
\begin{figure}[h]
\begin{center}
\mbox{\includegraphics[width=0.45\textwidth
]{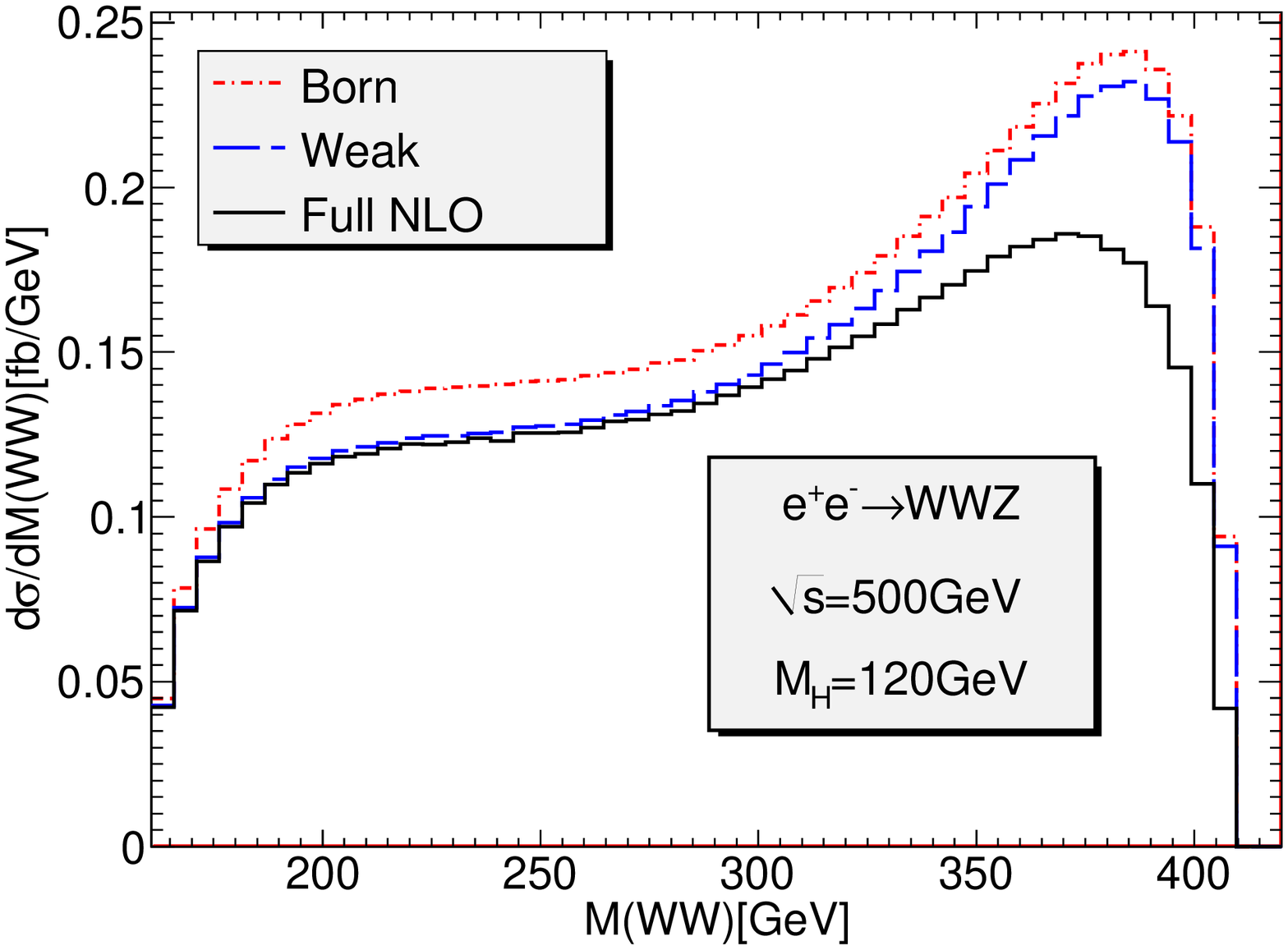} \hspace*{0.01\textwidth}
\includegraphics[width=0.45\textwidth]{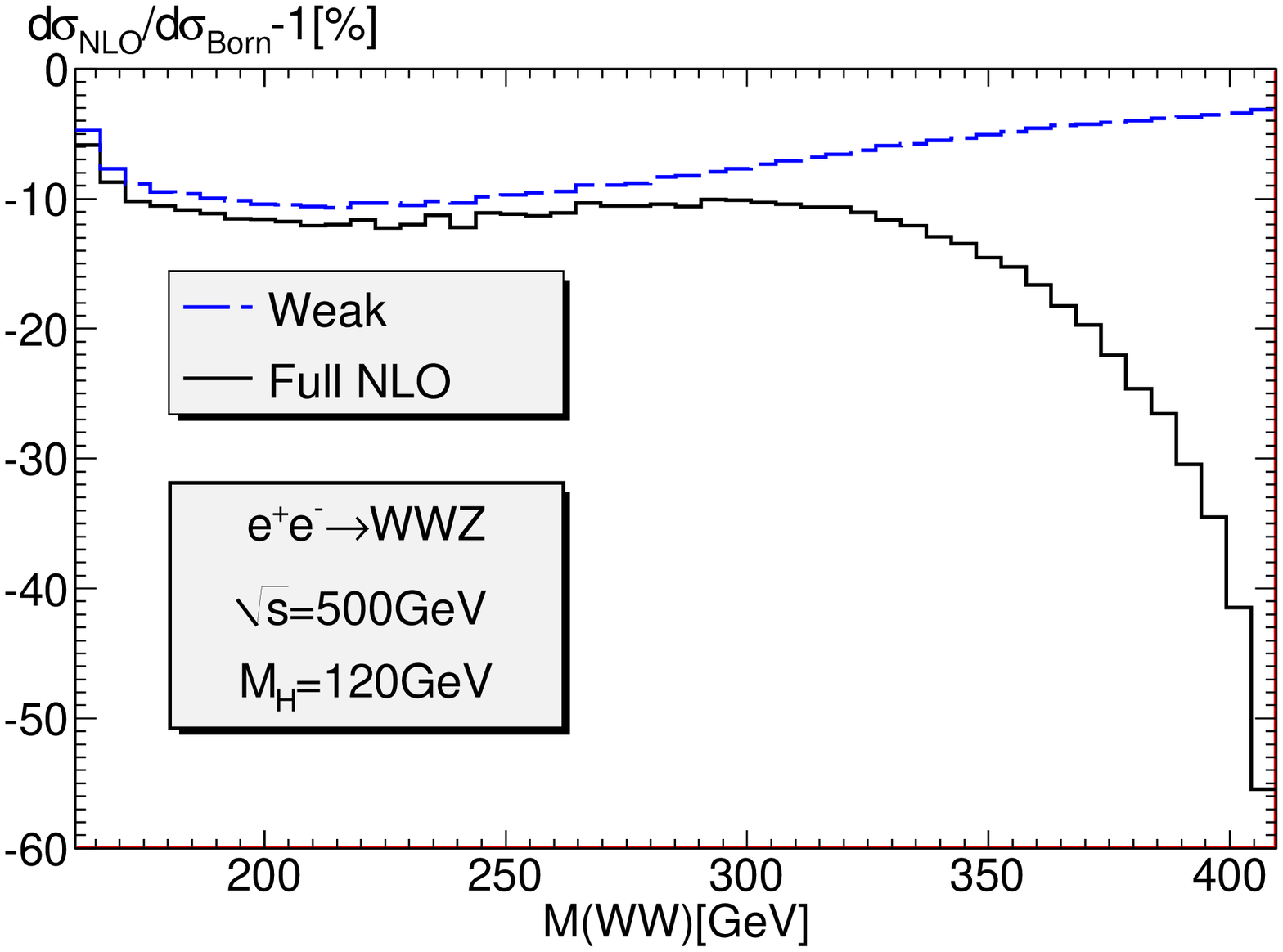}}
\mbox{\includegraphics[width=0.45\textwidth
]{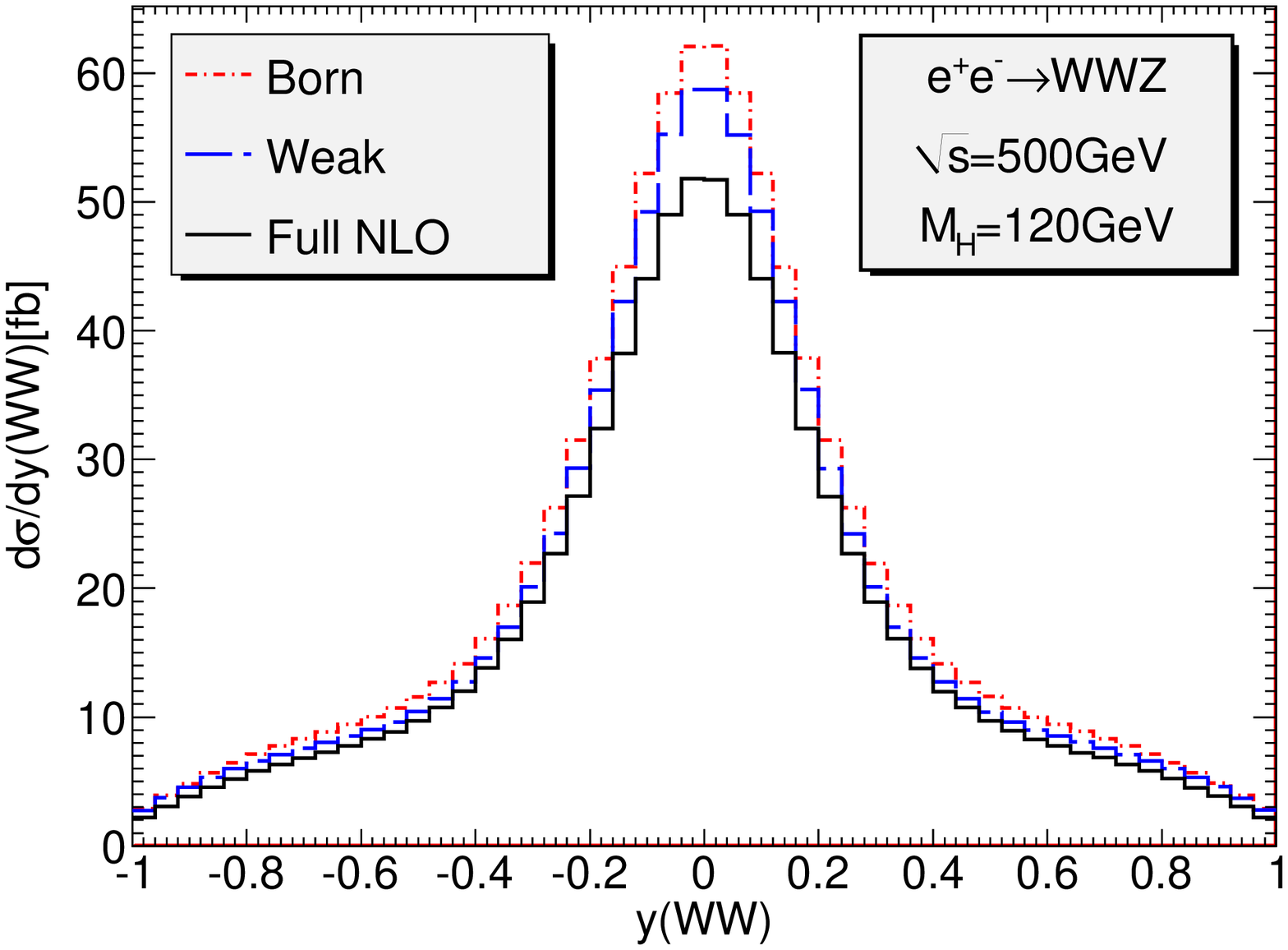} \hspace*{0.01\textwidth}
\includegraphics[width=0.45\textwidth]{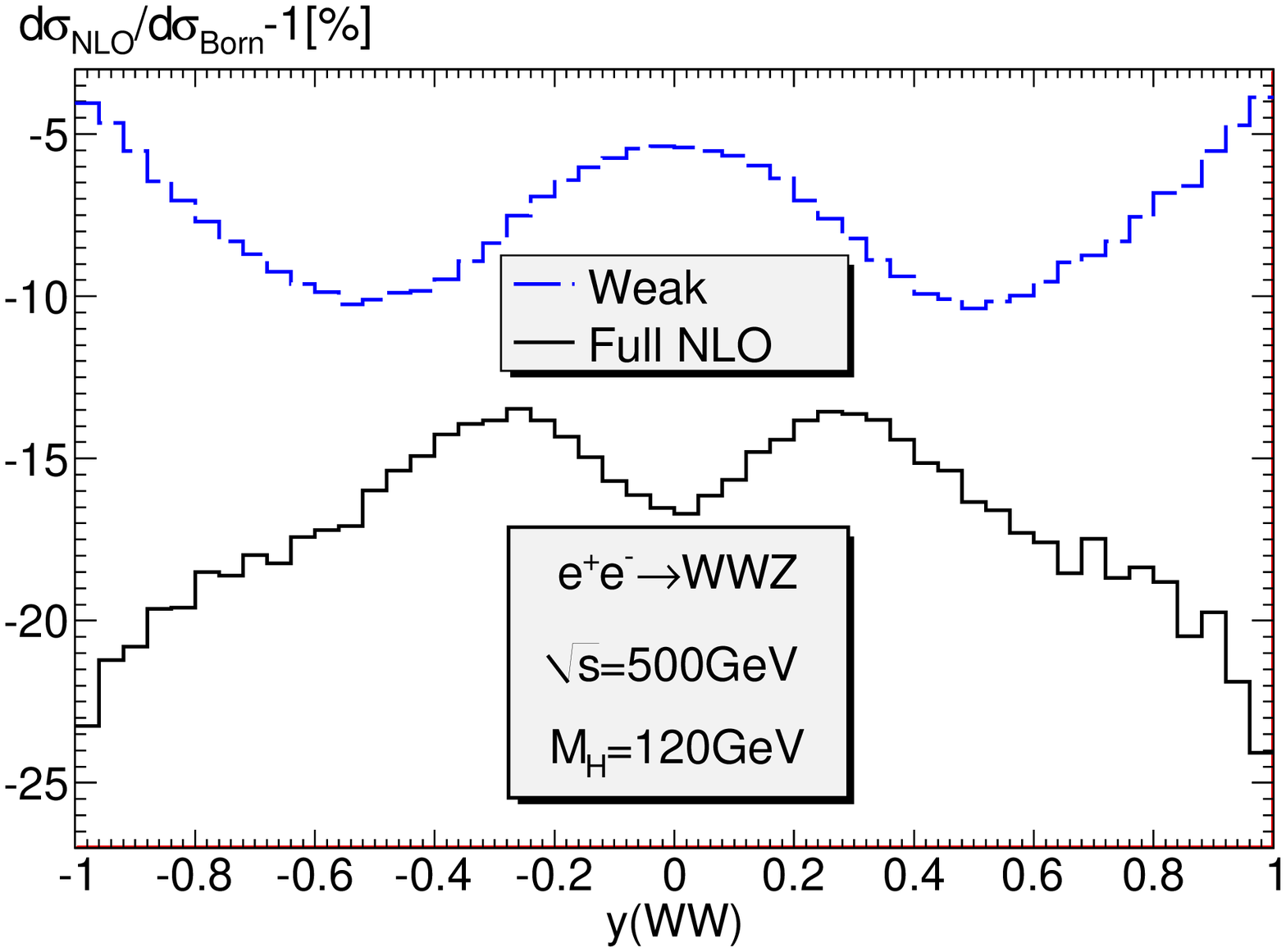}}
\caption{\label{ourwwzdist}{\em From top to bottom: distributions 
for the $WW$ invariant mass and the rapidity 
of the $WW$ system
for \eewwzt . 
The panels on
the left show the Born, the full NLO and the weak correction.
The panels on the right show the
corresponding relative (to the Born) percentage corrections.
}}
\end{center}
\end{figure}

\noindent {\underline{$\eewwz$:}}\\
Compared to $ZZZ$ production, the cross section for
\eewwzt is almost 2 orders of magnitudes larger for the same
centre-of-mass energy. For example at $500\gev$ it is about $40\fb$ at tree
level,
compared to $1\fb$ for the \eezzzt cross section. For an anticipated
 luminosity of $1 {\rm ab}^{-1}$, this means that
the cross section should be known at the
per-mil level. The behaviour of the total cross section as a function of
energy resembles that of \eezzztp. It rises sharply once the
threshold for production opens, reaches a peak before very slowly
decreasing as shown in Fig.~\ref{ourvvzsigma}. However as already
discussed the value of the peak is much larger, $\sim 50\fb$ at NLO,
moreover the peak is reached around $\sqrt{s}=1\tev$, much higher than
in $ZZZ$. This explains the bulk of the NLO corrections at lower
energies which are dominated by the QED correction, large and
negative around threshold and smaller at higher energies. As
the energy increases the weak corrections get larger
reaching about $-18\%$ at $\sqrt{s}=1.5\tev$. 
This is similar to the result of ZZZ production and is consistent with 
the behavior of double-logarithmic Sudakov corrections.   

In Fig.~\ref{ourwwzdist} we show the distributions in 
the $WW$ invariant mass and the rapidity 
of the $WW$ system.
Due to photon
radiation, in the full NLO corrections some large corrections do
show up at the edges of phase space. 
However, even after subtraction of the QED corrections the weak
corrections cannot be parameterized by an
overall scale factor, for all the distributions that we have
studied.

\section{Conclusions}
We have presented a calculation of the full next-to-leading order correction to the
processes \eewwzt and \eezzzt in the energy range of the
international linear collider and for Higgs masses below the $WW$
threshold. These processes would be the
successor of \eewwt in that they would measure the quartic
couplings $WWZZ$ and $ZZZZ$ which could retain residual effects of
the physics of electroweak symmetry breaking. With this in mind we
have subtracted the QED corrections and studied the genuine
weak  corrections in the $G_\mu$ scheme.
We find that the weak corrections can be
large and increase with the energy. 

\vspace{0.5cm}
\noi {\bf Acknowledgments} \\
LDN is grateful to the organisers of the Workshop, in particular Yoshimasa Kurihara, 
for their invitation and financial support. 


\end{document}